\documentclass[aps,prl,twocolumn,showpacs,superscriptaddress,groupedaddress,amsmath,amssymb]{revtex4-2}
\usepackage{graphicx}
\usepackage{latexsym}
\usepackage{bm} 
\usepackage{color}
\usepackage{epsfig}
\usepackage{multirow}
\usepackage{xcolor}
\usepackage{colortbl}
\usepackage{hhline}
\usepackage{simplewick}
\usepackage{soul}

\usepackage{graphicx}
\usepackage[colorlinks=true,linkcolor=blue,citecolor=blue,urlcolor=blue]{hyperref}
\usepackage{bbold}
\usepackage{gensymb}

\AtBeginDocument{%
    \newwrite\bibnotes
    \def\bibnotesext{Notes.bib}
    \immediate\openout\bibnotes=\jobname\bibnotesext
    \immediate\write\bibnotes{@CONTROL{REVTEX42Control}}
    \immediate\write\bibnotes{@CONTROL{%
    apsrev42Control,author="08",editor="1",pages="0",title="0",year="1"}}
     \if@filesw
     \immediate\write\@auxout{\string\citation{apsrev42Control}}%
    \fi
}%

\newcommand{\ket}[1]{|#1\rangle}
\newcommand{\mel}[3]{\langle#1|#2|#3\rangle}

\newcommand{\epvl}[1]{\langle#1\rangle}
\newcommand{\Epvl}[1]{\left\langle#1\right\rangle}
\def \Tr {\mathrm{Tr}}
\newcommand{\comment}[1]{}

\definecolor{mygreen}{rgb}{0, 0.7, 0}

\begin{document}

\title{Shiba duality and \(\eta\)-altermagnetism: Pairing and charge orders in bipartite attractive Hubbard models}

\author{Yu-Ping Lin}
\email{yuping.lin.056@gmail.com}
\altaffiliation[Present address: ]{IonQ, Inc., College Park, Maryland 20740, USA}
\affiliation{Department of Physics, University of California, Berkeley, California 94720, USA}

\date{\today}

\begin{abstract}
We show that Shiba duality maps the altermagnetic principle of momentum-dependent band splitting from spin to \(\eta\)-pseudospin, defining \(\eta\)-altermagnetism (\(\eta\)-ALM) as a Bogoliubov-de Gennes (BdG) counterpart of ALM in pairing and charge orders. In half-filled pure Hubbard models on bipartite lattices, the duality relates repulsion-driven antiferromagnetism (AFM) to attraction-driven \(\eta\)-AFM with uniform singlet pairing and staggered charge-density modulation. In the BdG bands, the Shiba-dual parity-time-reversal \(\mathcal{\tilde{P}}\mathcal{\tilde{T}}\) symmetry protects a Kramers degeneracy of \(\eta\)-pseudospin. Anisotropic second-neighbor hopping breaks this degeneracy and generates \(\eta\)-ALM. Odd-parity \(\eta\)-ALM shows \(\eta\)-pseudospin splitting, whereas even-parity \(\eta
\)-ALM has spin-\(\eta\)-locked splitting. Hartree-Fock-Bogoliubov computations on checkerboard and honeycomb lattices find \(\eta\)-ALMs with \(p\)-, \(d\)-, and \(f\)-wave splitting structures. Possible generalizations and experimental probes are discussed.
\end{abstract}

\maketitle

\addtocontents{toc}{\string\tocdepth@munge}

\textit{Introduction---} Altermagnetism (ALM) has emerged as a rapidly developing research direction in modern condensed matter physics \cite{hayami19jpsj,smejkal20sa,yuan20prb,ma21nc,mazin21pnas,smejkal22prx1,smejkal22prx2, jungwirth25nt, lin25ax, lee24prl, reimers24nc, osumi24prb, ding24prl, zhou25n}. As the third class of collinear magnetism in modern classification, ALM goes beyond ferromagnetism (FM) and antiferromagnetism (AFM) by combining nonrelativistic spin splitting and zero net magnetization. With different symmetry breakings in the underlying crystal, the spin splitting in ALM can take different symmetry structures in momentum space. If a space-group symmetry is broken, the ALM can have an even-parity form factor \cite{smejkal22prx1, smejkal22prx2, roig24prb, durrnagel25prl}. Meanwhile, time-reversal symmetry breaking can lead to odd-parity ALMs \cite{lin25ax, huang26prl, zhu26prl, li26prl, liu26prb, leeb26ax}. The unconventional structure of ALM supports many unique applications, including spintronics development \cite{bai24afm, duan25prl, gu25prl} and intricate interplay with superconductivity \cite{mazin25aap, zhu23prb, brekke23prb, giil24prb, chakraborty24prb, bose24prb, kim25prb, wu25prl, papaj23prb, sun23prb, ouassou23prl, beenakker23prb}.

The discovery of ALM belongs to the long exploration of higher-angular-momentum states in superconducting \cite{sigrist91rmp, tsuei00rmp, nandkishore12np}, charge \cite{pomeranchuk59sp, kivelson98n, nayak00prb, halboth00prl, chakravarty01prb, oganesyan01prb, venderbos16prb1, lin19prb, lin21prb}, and spin \cite{pomeranchuk59sp, hirsch90prb, gorkov92prl, wu04prl, wu07prb, venderbos16prb2, kiselev17prb, wu18prb, classen20prb, birkhellenes23ax, luo26ax, mitscherling26ax} orders. An initial motivation was the search for robust \(d\)-wave magnetism as a counterpart of \(d\)-wave superconductivity \cite{smejkal22prx2}. The rapid development of ALM has further motivated the search for band splitting in other orders, including orbital ALM \cite{pan25ax, li26ax} and alterelectricity \cite{fang26ax}. ALM has an important distinction from previously studied higher-angular-momentum ordered states. In previously studied states, higher angular momenta are induced by symmetry-breaking orders in the corresponding channels. Many of these orders develop on bonds instead of sites, especially in superconductivity, where onsite pairing orders are usually \(s\)-wave. Meanwhile, ALM develops conventional onsite spin orders and acquires higher angular momentum from the underlying crystal. From this perspective, the magnetic counterpart of \(d\)-wave superconductivity is \(d\)-wave spin-bond order, not \(d\)-wave ALM. This comparison raises an important question: {\it What is the natural counterpart of ALM in superconductivity?}

In this {\it Letter}, we propose the concept of \(\eta\)-ALM as an answer to this question. Our study focuses on the particle-hole symmetric Hubbard models at half filling on general 2D and 3D bipartite lattices. An \(\text{SU}(2)\) symmetry implies the definition of \(\eta\)-pseudospin \cite{yang89prl, yang90mpl, zhang91ijmpb}, whose components correspond to onsite pairing and charge orders. Importantly, the Shiba duality establishes an exact correspondence between the repulsive and attractive models \cite{shiba72ptp, emery76prb}. In the pure Hubbard model, this duality implies the correspondence of repulsion-driven AFM and attraction-driven \(\eta\)-AFM ground states \cite{robaszkiewicz81prb01,robaszkiewicz81prb02,scalettar89prl, duchon13ax}. Revisiting the Shiba duality, we derive the Kramers degeneracy of \(\eta\)-pseudospins in the Bogoliubov-de Gennes (BdG) bands in \(\eta\)-AFM. This derivation establishes the general correspondence of symmetry and band-structure properties in spin and \(\eta\)-pseudospin orders. With anisotropic second-neighbor hopping, repulsion-driven AFM becomes even- and odd-parity ALMs. Our duality framework naturally finds the corresponding \(\eta\)-ALMs under attraction. Odd-parity \(\eta\)-ALM manifests \(\eta\)-pseudospin splitting in the BdG bands. Meanwhile, even-parity \(\eta\)-ALM shows spin-\(\eta\)-locked splitting. By Hartree-Fock-Bogoliubov computations, we present \(\eta\)-ALMs on 2D checkerboard and honeycomb lattices.

We emphasize that \(\eta\)-ALMs are not conventional higher-angular-momentum ordered states. Instead, they are BdG manifestations of the altermagnetic principle in pairing and charge orders: While the orders remain onsite, higher-angular momentum and band splitting originate from the underlying crystal. This mechanism is conceptually analogous to proximity-induced topological superconductivity \cite{fu08prl}. However, an important difference is that the nontrivial BdG structure in \(\eta\)-ALM is primarily \(\eta\)-altermagnetic rather than topological. Our work opens a unique route to the search for higher-angular-momentum ordered states in unconventional phases of matter.

\textit{Bipartite pure Hubbard model---}We begin with the pure Hubbard model on general 2D and 3D bipartite lattices with two-site unit cells
\begin{equation}
    \label{eq:hubbard}
    H = -t_1\sum_{\epvl{ij}_1}c_i^\dagger c_j + U_0\sum_in_{i\uparrow}n_{i\downarrow} - \mu\sum_in_i.
\end{equation}
Here \(c_i^{(\dagger)} = [(c_{i\uparrow}, c_{i\downarrow})^T]^{(\dagger)}\) annihilates (creates) fermions at site \(i\) with spins \(\sigma=\uparrow,\downarrow\), while \(n_i = c_i^\dagger c_i = \sum_\sigma n_{i\sigma}\) and \(n_{i\sigma} = c_{i\sigma}^\dagger c_{i\sigma}\) measure the total and spin-\(\sigma\) densities at site \(i\). The model contains the hopping \(t_1\) between first neighbors \(\epvl{ij}_1\) and the onsite interaction \(U_0\). The chemical potential \(\mu = U_0 / 2\) is chosen, which sets the site-averaged fermion density at half filling \(n_f = \overline{\epvl{n_i}} = 1\) in the ground states. Notably, the Hamiltonian (\ref{eq:hubbard}) at this filling has a particle-hole symmetry \(\mathcal{C}\): \(c_i \rightarrow (-1)^ic_i^\dagger\), \(n_i \rightarrow 2 - n_i\) \cite{chiu16rmp}, where we define \(i \equiv 0, 1 \text{ (mod \(2\))}\) on the two sublattices.

\textit{Shiba duality, AFM, and \(\eta\)-AFM---}Another relevant transformation in the model (\ref{eq:hubbard}) is the Shiba transformation \cite{shiba72ptp, emery76prb}. This partial particle-hole transformation
\begin{equation}
    c_{i\uparrow} \rightarrow c_{i\uparrow}, \quad
    c_{i\downarrow} \rightarrow (-1)^ic_{i\downarrow}^\dagger
\end{equation}
maps a repulsive model with \(U_0 = U > 0\) exactly to its attractive counterpart with \(U_0 = -U < 0\). Therefore, the Shiba duality establishes an exact correspondence between the repulsive and attractive models.

An important consequence of Shiba duality is the exact correspondence between the ground states \cite{robaszkiewicz81prb01,robaszkiewicz81prb02,scalettar89prl, duchon13ax}. For the repulsive model, the ground state develops AFM under sufficiently strong repulsion \(U > U_c\). The critical interaction \(U_c\) is lattice-dependent, which vanishes \(U_c = 0\) under nonzero density of states \cite{hirsch85prb, hirsch89prl, varney09prb, duchon13ax} and is nonzero \(U_c > 0\) otherwise \cite{sorella92epl,assaad13prx,otsuka16prx, zhang09prb}. The AFM exhibits staggered onsite spin orders on the two sublattices
\begin{equation}
    \mathbf{s}_i = \epvl{\mathbf{S}_i} = (-1)^i\mathbf{m}, \quad 
    \mathbf{S}_i = c_i^\dagger \left(\frac{\boldsymbol{\sigma}}{2}\right) c_i.
\end{equation}
Here the spin order \(\mathbf{m} = m \mathbf{\hat m}\) has magnitude \(m > 0\) along a normal vector \(\mathbf{\hat m}\), \(\boldsymbol{\sigma} = (\sigma^1, \sigma^2, \sigma^3)\) are the spin Pauli matrices, and \(\epvl{\dots}\) is the ground-state expectation value. AFM breaks the global \(\text{SU}^s(2)\) symmetry of the spin \(\mathbf{S} = \sum_i\mathbf{S}_i\) with degeneracy among different directions \(\mathbf{\hat m}\). On the other hand, the ground state in the attractive model can be inferred from the Shiba duality \cite{robaszkiewicz81prb01,robaszkiewicz81prb02,scalettar89prl,duchon13ax}. Under the Shiba transformation, the spin \(\mathbf{S}_i\) transforms into the \(\eta\)-pseudospin
\begin{equation}
    \mathbf{\tilde{S}}_i = \frac{1}{2}\psi_i^\dagger \left(\frac{\boldsymbol{\eta}_i}{2}\right) \psi_i.
\end{equation}
The Nambu spinor \(\psi_i = (c_i, (i\sigma^2)(c_i^\dagger)^T)^T\) is defined with the time-reversal spin flip \(i\sigma^2\). Meanwhile, \(\boldsymbol{\eta}_i = ((-1)^i\rho^1, (-1)^i\rho^2, \rho^3)\) are the \(\eta\)-pseudospin Pauli matrices with particle-hole Pauli matrices \(\boldsymbol{\rho} = (\rho^1, \rho^2, \rho^3)\) in the Nambu space. The model has a global \(\text{SU}^\eta(2)\) symmetry of the \(\eta\)-pseudospin \(\mathbf{\tilde{S}} = \sum_i\mathbf{\tilde{S}}_i\) \cite{yang89prl,yang90mpl,zhang91ijmpb}. Under sufficiently strong attraction \(U_0 < -U_c\), the ground state develops \(\eta\)-AFM
\begin{equation}
    \mathbf{\tilde{s}}_i = \epvl{\mathbf{\tilde{S}}_i} = (-1)^i\mathbf{m}.
\end{equation}
This order breaks the \(\text{SU}^\eta(2)\) symmetry with degeneracy among different directions \(\mathbf{\hat m}\). Note that the components correspond to onsite uniform singlet pairing (USP) \((-1)^i\tilde{s}_i^{1,2} = \text{Re/Im} (\epvl{c_{i\uparrow}^\dagger c_{i\downarrow}^\dagger})\) and staggered charge-density-modulation (CDM) \(\tilde{s}_i^3 = (-1)^i(\epvl{n_i} - 1)/2\) orders, which are degenerate in the ground state \cite{robaszkiewicz81prb01,robaszkiewicz81prb02,scalettar89prl,duchon13ax}.

\textit{Spin and \(\eta\)-pseudospin band structures---}The spin band structure is an important pillar in the modern classification of magnetism \cite{smejkal22prx1, smejkal22prx2}. Meanwhile, the correspondence to \(\eta\)-pseudospin band structure is much less explored. Here we discuss the band-structure correspondence between AFM and \(\eta\)-AFM, which is an important basis for our later duality study of ALM and \(\eta\)-ALM.

At the mean-field level, the repulsive model (\ref{eq:hubbard}) has an effectively noninteracting approximation in AFM
\begin{equation}
    \label{eq:ham_afm}
    H^\text{AFM} = -t_1\sum_{\epvl{ij}_1}c_i^\dagger c_j - \sum_i2\mathbf{s}_i \cdot \mathbf{S}_i.
\end{equation}
This model obeys the parity-time-reversal \(\mathcal{P} \mathcal{T}\) symmetry, where the parity \(\mathcal{P}\): \(c_i \rightarrow c_{-i}\) about a first-neighbor bond center switches the sublattices and the time reversal \(\mathcal{T}\): \(c_i \rightarrow i\sigma^2 c_i\) with complex conjugation obeys \(\mathcal{T}^2 = -1\). To study the band structure, we consider the momentum-space Hamiltonian \(H^\text{AFM} = \sum_{\mathbf{k}}c_{\mathbf{k}}^\dagger \mathcal{H}^\text{AFM}_{\mathbf{k}} c_{\mathbf{k}}\). Here \(c_{\mathbf{k}}^{(\dagger)} = [(c_{\mathbf{k}\tau\sigma})^T]^{(\dagger)}\) annihilates (creates) fermions at momentum \(\mathbf{k}\) on sublattices \(\tau = 0, 1\) with spins \(\sigma = \uparrow, \downarrow\). The Bloch Hamiltonian takes the form
\begin{equation}
    \label{eq:bloch_ham_afm}
    \mathcal{H}^\text{AFM}_{\mathbf{k}} = h_{1\mathbf{k}}\tau^1 + h_{2\mathbf{k}}\tau^2 - \mathbf{m} \cdot \tau^3 \boldsymbol{\sigma}
\end{equation}
with the sublattice Pauli matrices \(\boldsymbol{\tau} = (\tau^1, \tau^2, \tau^3)\). The real functions \(h_{1,2\mathbf{k}} = \pm h_{1,2(-\mathbf{k})}\) correspond to the first-neighbor hopping \(h_{1\mathbf{k}} - ih_{2\mathbf{k}} = -t_1\sum_{l = 1}^{N_1}e^{-i \mathbf{k} \cdot \mathbf{a}_{1l}}\), where \(\mathbf{a}_{1l}\) are all \(N_1\) first-neighbor vectors from \(\tau = 1\) to \(0\). Note that the Bloch Hamiltonian obeys the \(\mathcal{P} \mathcal{T}\) symmetry \((PT)\mathcal{H}^\text{AFM}_{\mathbf{k}}(PT)^{-1} = \mathcal{H}^\text{AFM}_{\mathbf{k}}\), where \(P = \tau^1\) and \(T = i\sigma^2K\) with complex conjugation \(K\). Since the \(\mathcal{P} \mathcal{T}\) symmetry protects the Kramers degeneracy, the bands have double degeneracy of spins with dispersion energies
\begin{equation}
    \label{eq:energy_afm}
    E^\pm_{\mathbf{k}} = \pm\sqrt{h_{1\mathbf{k}}^2 + h_{2\mathbf{k}}^2 + m^2}.
\end{equation}

For attraction-driven \(\eta\)-AFM, the \(\eta\)-pseudospin band structure can be studied by a similar approach. The mean-field model takes the form (\ref{eq:hubbard})
\begin{equation}
    \label{eq:mfham_etaafm}
    H^{\eta\text{-AFM}} = \frac{1}{2}\sum_{\epvl{ij}_1}\psi_i^\dagger (-t_1\rho^3) \psi_j - \sum_i 2 \mathbf{\tilde{s}}_i \cdot \mathbf{\tilde{S}}_i
\end{equation}
and obeys the Shiba-dual \(\mathcal{\tilde{P}} \mathcal{\tilde{T}}\) symmetry. Here the Shiba-dual symmetries are defined as \(\mathcal{\tilde{P}}\): \(\psi_i \rightarrow \rho^3 \sigma^3 \psi_{-i}\) and \(\mathcal{\tilde{T}}\): \(\psi_i \rightarrow (-1)^i i\rho^2 \psi\) with complex conjugation. The momentum-space Hamiltonian reads \(H^{\eta\text{-AFM}} = (1/2)\sum_{\mathbf{k}}\psi_{\mathbf{k}}^\dagger \mathcal{H}^{\eta\text{-AFM}}_{\mathbf{k}} \psi_{\mathbf{k}}\), where \(\psi_{\mathbf{k}} = (c_{\mathbf{k}}, (i\sigma^2)(c_{-\mathbf{k}}^\dagger)^T)^T\) is the momentum-space Nambu spinor. We have the BdG Hamiltonian
\begin{equation}
    \label{eq:bdg_ham_etaafm}
    \mathcal{H}^{\eta\text{-AFM}}_{\mathbf{k}} = \rho^3(h_{1\mathbf{k}}\tau^1 + h_{2\mathbf{k}}\tau^2) - \mathbf{m} \cdot \tau^3\boldsymbol{\eta}
\end{equation}
with the \(\eta\)-pseudospin Pauli matrices \(\boldsymbol{\eta} = (\rho^1\tau^3, \rho^2\tau^3, \rho^3)\). The BdG bands are doubly degenerate in spins. Meanwhile, the Shiba-dual \(\mathcal{\tilde{P}} \mathcal{\tilde{T}}\) symmetry \((\tilde{P}\tilde{T})\mathcal{H}^{\eta\text{-AFM}}_{\mathbf{k}}(\tilde{P}\tilde{T})^{-1} = \mathcal{H}^{\eta\text{-AFM}}_{\mathbf{k}}\) protects the Kramers degeneracy of \(\eta\)-pseudospins, where \(\tilde{P} = \rho^3\tau^1\sigma^3\) and \(\tilde{T} = i\rho^2\tau^3K\). Indeed, the BdG Hamiltonian gives an identical band structure (\ref{eq:energy_afm}) to the AFM with four-fold degenerate bands. This result confirms an exact correspondence between the spin and \(\eta\)-pseudospin band structures under the Shiba duality.

Note that the Shiba duality applies to general spin and \(\eta\)-pseudospin orders. For example, FM is dual to \(\eta\)-FM with staggered onsite \(\eta\)-pairing \cite{yang89prl} and uniform charge-density orders. While FM can occur away from half filling under strong repulsion \cite{nagaoka66prl}, \(\eta\)-pairing order may appear similarly under strong attraction \cite{singh91prl}. The spin and \(\eta\)-pseudospin splittings share the same \(s\)-wave structure in momentum space.

\textit{ALMs in repulsive model---}Having seen the Kramers degeneracies in AFM and \(\eta\)-AFM, we discuss how they can be broken in the same orders. On the magnetism side, breaking the \(\mathcal{P}\mathcal{T}\) symmetry can lead to ALM with alternate spin splitting around the Brillouin-zone (BZ) center \(\Gamma\) \cite{jungwirth25nt, lin25ax}. Meanwhile, the corresponding \(\eta\)-ALM is unknown. Here we introduce a set of ALMs and their spin band structures, which are important hints to our discovery of \(\eta\)-ALM.

An essential element in ALM is an anisotropic sublattice-dependent term \cite{jungwirth25nt, lin25ax}. This term breaks the \(\mathcal{P}\mathcal{T}\) symmetry, lifts the sublattice degeneracy \cite{lin24axsp}, and reduces the space group \(\mathbf{G}\) to a subgroup \(\mathbf{G'}\). The resulting ALM preserves the spin-group symmetry \([E^s || \mathbf{H'}] + [C^s_2 || \mathbf{G'} - \mathbf{H'}]\) \cite{jungwirth25nt} that protects the compensated collinear spin order. The first and second entities in the brackets are the spin and space-time symmetries, respectively. \(E^s\) is the spin identity, \(C^s_2\) is a twofold rotation orthogonal to the collinear spin direction, and \(\mathbf{H'}\) is the half subgroup of \(\mathbf{G'}\) in the staggered spin order. We further impose the particle-hole symmetry \(\mathcal{C}\) so that the Shiba duality retains a clean form. The simplest term is the anisotropic second-neighbor hopping
\begin{equation}
    \label{eq:2nb}
    \delta H^2 = -t_2\sum_{\epvl{ij}_2} c_i^\dagger \nu_{ij} c_j
\end{equation}
with anisotropy configuration \(\{\nu_{ij}\}\). We consider the configurations \(\{\nu_{ij}\}\) that keep \(\mathbf{G}'\) halvable in the staggered spin order. Note that these bonds are first-neighbor on the sublattices, which form a pair of parallel Bravais lattices. We assign uniform hopping along each Bravais-lattice axis with opposite signs on the two sublattices.

In ALM, the mean-field Hamiltonian (\ref{eq:ham_afm}) becomes \(H^\text{ALM} = H^\text{AFM} + \delta H^2\) with the Bloch Hamiltonian (\(\ref{eq:bloch_ham_afm}\))
\begin{equation}
    \mathcal{H}^\text{ALM}_{\mathbf{k}} = \mathcal{H}^\text{AFM}_{\mathbf{k}} + h_{3\mathbf{k}}\tau^3.
\end{equation}
The additional term with real function \(h_{3\mathbf{k}}\) breaks the \(\mathcal{P} \mathcal{T}\) symmetry. Since the Kramers degeneracy is not protected, the bands can develop spin splitting
\begin{equation}
    \label{eq:energy_alm}
    E^\pm_{\mathbf{k} \sigma^m} = \pm\sqrt{h_{1\mathbf{k}}^2 + h_{2\mathbf{k}}^2 + (h_{3\mathbf{k}} - \sigma^m m)^2}.
\end{equation}
The spin component \(S^m = \mathbf{\hat m} \cdot \mathbf{S}\) remains a good quantum number \([H^\text{ALM}, S^m] = 0\). Correspondingly, the bands with Bloch states \(\ket{u_{\mathbf{k}}}\) have well-defined spin indices \(\sigma^m = \mel{u_{\mathbf{k}}}{\mathbf{\hat{m}} \cdot \boldsymbol{\sigma}}{u_{\mathbf{k}}} = \pm 1\).
Under the spin-group symmetry \([C^s_2 || \mathbf{G}' - \mathbf{H}']\), the band structure obeys the symmetry relation \(E^\pm_{\mathbf{k} \sigma^m} = E^\pm_{(g\mathbf{k})(-\sigma^m)}\) for \(g \in \mathbf{G}' - \mathbf{H}'\) and has alternate spin splitting around \(\Gamma\).

The parity of ALM is determined by the anisotropic second-neighbor hopping. For even parity \cite{jungwirth25nt}, we have sublattice bonds with real hopping \(\nu_{ij} = 0, \pm1\). The positive and negative bonds on the same sublattice are related by \(g \in \mathbf{G}' - \mathbf{H}'\) and are exchanged on the other sublattice. In momentum space, the even-parity spin splitting \(E^\pm_{\mathbf{k}\sigma^m} = E^\pm_{(-\mathbf{k})\sigma^m}\) is induced by the even function \(h^\text{even}_{3\mathbf{k}} = -2t_2\sum_{l=1}^{N_{2l}/2}\nu_l\cos(\mathbf{k} \cdot \mathbf{a}_{2l})\) under the spin-group symmetry \([C^s_2\mathcal{T}||\mathcal{T}]\). Here \(\mathbf{a}_{2l}\) are the \(N_{2l} / 2\) nonparallel Bravais-lattice vectors and \(\nu_l=0,\pm1\). On the other hand, odd-parity ALM \cite{lin25ax} occurs under sublattice currents with imaginary hopping \(\nu_{ij} = 0, \pm i\). The odd function \(h^\text{odd}_{3\mathbf{k}} = 2t_2\sum_{l=1}^{N_{2l}/2}\nu_l\sin(\mathbf{k}\cdot\mathbf{a}_{2l})\) with \(\nu_l = 0, \pm 1\) leads to odd-parity spin splitting \(E^\pm_{\mathbf{k}\sigma^m} = E^\pm_{(-\mathbf{k})(-\sigma^m)}\) in the band structure. Interestingly, the compensated collinear spin order is always protected by the spin-group symmetry \([C^s_2||\mathcal{P}]\), allowing robust ALMs under broader anisotropy configurations \(\{\nu_{ij}\}\).

\textit{\(\eta\)-ALMs in attractive model---}We now introduce our discovery of analogous \(\eta\)-pseudospin splitting in \(\eta\)-ALMs. Similar to \(\eta\)-AFM, we obtain \(\eta\)-ALMs by applying the Shiba transformation to ALMs. An important distinction is the noninvariance of second-neighbor hopping \(\delta H^2 \rightarrow \delta \tilde{H}^2\). While imaginary hopping remains invariant \(\nu_{ij} \rightarrow \tilde{\nu}_{ij} = \nu_{ij}\), real hopping becomes spin-dependent \(\nu_{ij} \rightarrow \tilde{\nu}_{ij} = \nu_{ij}\sigma^3\). Therefore, the mean-field Hamiltonians \(H^{\eta\text{-ALM}} = H^{\eta\text{-AFM}} + \delta \tilde{H}^2\) in \(\eta\)-ALMs have the BdG Hamiltonians (\ref{eq:bdg_ham_etaafm})
\begin{equation}
    \label{eq:bdg_ham_etaalm}
    \mathcal{H}^{\eta\text{-ALM}}_{\mathbf{k}} = \mathcal{H}^{\eta\text{-AFM}}_{\mathbf{k}} + \left\{\begin{aligned}
    &h^\text{odd}_{3\mathbf{k}}\tau^3, && \text{odd-parity}, \\
    &h^\text{even}_{3\mathbf{k}}\tau^3\sigma^3, && \text{even-parity}.
    \end{aligned}\right.
\end{equation}
\(\eta\)-ALMs preserve the \(\eta\)-pseudospin-group symmetry \([E^\eta || \mathbf{\tilde{H}}'] + [C^\eta_2 || \mathbf{\tilde{G}'} - \mathbf{\tilde{H}'}]\) that protects the compensated collinear \(\eta\)-pseudospin order. Here \(E^\eta\) and \(C^\eta_2\) are analogous to their spin counterparts, while \(\mathbf{\tilde{G}'}\) and \(\mathbf{\tilde{H}'}\) are the Shiba-dual counterparts of \(\mathbf{G'}\) and \(\mathbf{H'}\).

Since the Shiba-dual \(\mathcal{\tilde{P}} \mathcal{\tilde{T}}\) symmetry is broken, the Kramers degeneracy of \(\eta\)-pseudospins is not protected. The BdG band structure develops band splitting with the same dispersion energies (\ref{eq:energy_alm}) as in ALM. The \(\eta\)-pseudospin component \(\tilde{S}^m = \mathbf{\hat{m}} \cdot \mathbf{\tilde{S}}\) remains a good quantum number \([H^{\eta\text{-ALM}}, \tilde{S}^m] = 0\). Correspondingly, the bands with BdG states \(\ket{u_{\mathbf{k}}}\) can be labeled by the \(\eta\)-pseudospin indices \(\eta^m = \mel{u_{\mathbf{k}}}{\mathbf{\hat{m}} \cdot \boldsymbol{\eta}}{u_{\mathbf{k}}} = \pm 1\). Under the \(\eta\)-pseudospin-group symmetry \([C^\eta_2 || \mathbf{\tilde{G}'} - \mathbf{\tilde{H}'}]\), the bands obey the symmetry relation \(E^\pm_{\mathbf{k}\eta^m} = E^\pm_{(\tilde{g}\mathbf{k})(-\eta^m)}\) for \(\tilde{g} \in \mathbf{\tilde{G}'} - \mathbf{\tilde{H}'}\). However, the band splitting has different labels in odd- and even-parity \(\eta\)-ALMs. This distinction originates from the difference in the second-neighbor hoppings (\ref{eq:bdg_ham_etaalm}).

For odd parity, the Shiba duality of ALM and \(\eta\)-ALM is clean. The band splitting in \(\eta\)-ALM has the same form as in ALM (\ref{eq:energy_alm})
\begin{equation}
    E^{\text{odd},\pm}_{\mathbf{k}\eta^m} = E^{\text{odd},\pm}_{\mathbf{k}(\sigma^m\rightarrow\eta^m)}
\end{equation}
under a direct index change. The odd-parity splitting \(E^\pm_{\mathbf{k}\eta^m} = E^\pm_{(-\mathbf{k})(-\eta^m)}\) occurs under the \(\eta\)-pseudospin-group symmetry \([C^\eta_2 || \mathcal{\tilde{P}}]\), which protects the compensated collinear \(\eta\)-pseudospin order.

The even-parity duality is less direct due to spin-dependent second-neighbor hopping (\ref{eq:bdg_ham_etaalm}). Since the spin component \(S^3\) is a good quantum number \([H^{\eta\text{-ALM}}, S^3] = 0\), the bands can be labeled by the spin indices \(\sigma^3 = \mel{u_{\mathbf{k}}}{\sigma^3}{u_{\mathbf{k}}} = \pm 1\). Interestingly, the dispersion energies
\begin{equation}
    E^{\text{even},\pm}_{\mathbf{k}\eta^m\sigma^3} = E^{\text{even},\pm}_{\mathbf{k}(\sigma^m\rightarrow\eta^m\sigma^3)}
\end{equation}
carry the spin-\(\eta\)-locked indices \(\eta^m\sigma^3 = \pm 1\). In each spin branch \(\sigma^3 = \pm 1\), the bands obey the even-parity relation \(E^\pm_{\mathbf{k}\eta^m\sigma^3} = E^\pm_{-\mathbf{k}\eta^m\sigma^3}\) under the \(\eta\)-pseudospin-group symmetry \([C^\eta_2 \mathcal{\tilde{T}} || \mathcal{\tilde{T}}]\). However, the \(\eta\)-pseudospin splitting is opposite in the two spin branches \(\sigma^3 = \pm 1\), leading to the spin-\(\eta\) locking \(E^\pm_{\mathbf{k}\eta^m\sigma^3} = E^\pm_{\mathbf{k}(-\eta^m)(-\sigma^3)}\). Note that under even parity, spin-independent second-neighbor hopping does not induce band splitting. The reason is that the mass term \(h^\text{even}_{3\mathbf{k}}\rho^3\tau^3 - \mathbf{m} \cdot \tau^3\boldsymbol{\eta}\) gives a single correction \([m_1^2 + m_2^2 + (h^\text{even}_{3\mathbf{k}} - m_3)^2]\) to the dispersion energies (\ref{eq:energy_afm}) of \(\eta\)-AFM. Therefore, even-parity \(\eta\)-ALM generally manifests  spin-\(\eta\)-locked splitting in the bands.

\textit{2D-lattice examples---}\(\eta\)-ALMs can occur broadly in the ground states on general 2D and 3D bipartite lattices \cite{lin25ax}. As concrete examples, we present \(\eta\)-ALMs on the 2D checkerboard and honeycomb lattices. Our demonstration is based on the Hartree-Fock-Bogoliubov theory \cite{supp}, which computes the mean-field ground states of interacting fermions with charge, spin, and pairing orders. The computation is applied to the models with \(t_1 = 1\), \(t_2 = 0.1\), \(U_0 = -4\), and \(\mu = U_0 / 2 = -2\) on the finite checkerboard (\(16^2\times2\)) and honeycomb (\(18^2\times2\)) lattices with periodic boundary conditions. Note that odd-parity \(\eta\)-ALMs can be compared with odd-parity ALMs in previous study \cite{lin25ax}.

\begin{figure}[t]
\centering
\includegraphics[scale=1]{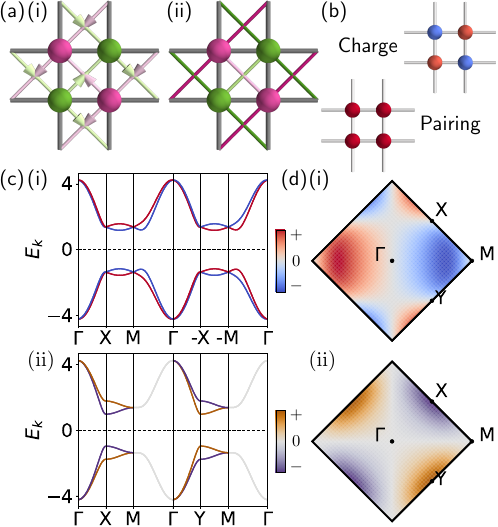}
\caption{\label{fig:checkerboard} \(\eta\)-ALMs on the checkerboard lattice. (a) The lattice with two sublattices (green and pink). (i) The odd-parity model has sublattice currents and (ii) the even-parity model has sublattice spin bonds (darker and lighter for \(\pm\sigma^3\) and blank for \(0\)). (b) CDM and USP orders in the ground state, where we plot \((\epvl{n_i} - 1) / 2\) and \(|\epvl{c_{i\uparrow}^\dagger c_{i\downarrow}^\dagger}|\text{sign}(\text{Re}(\epvl{c_{i\uparrow}^\dagger c_{i\downarrow}^\dagger}))\), respectively. (c) The band structures in \(\eta\)-ALMs have (i) odd-parity \(\eta\)-pseudospin splitting with \(\eta^m = \pm 1\) and (ii) even-parity spin-\(\eta\)-locked splitting with \(\eta^m \sigma^3 = \pm 1\). (d) The BZ mapping of splitting energy show the (i) \(p\)-wave splitting in odd-parity \(\eta\)-ALM and (ii) \(d\)-wave splitting in even-parity \(\eta\)-ALM. The colors in (b)-(d) represent the respective data and follow the respective color bars.}
\end{figure}

\begin{figure}[t]
\centering
\includegraphics[scale=1]{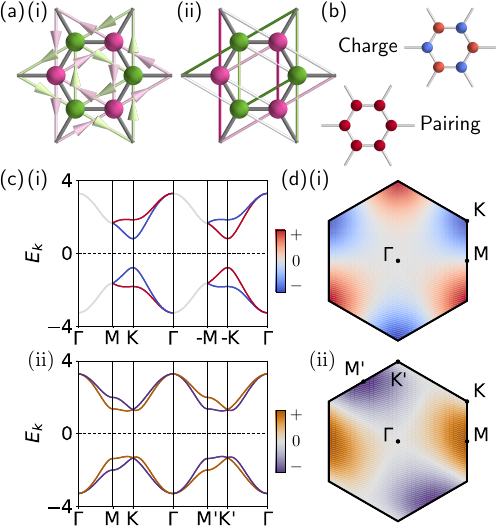}
\caption{\label{fig:honeycomb} \(\eta\)-ALMs on the honeycomb lattice. For the (i) odd- and (ii) even-parity models, we show (a) the lattice model, (b) the ground-state CDM and USP orders, (c) the splitting in the bands, and (d) the splitting energies with (i) \(f\)-wave and (ii) \(d\)-wave structures. The plotting formats follow Fig.~\ref{fig:checkerboard}.}
\end{figure}

We first consider the checkerboard lattice (Fig.~\ref{fig:checkerboard}). In the odd-parity model, we introduce the sublattice currents with \(\tilde{\nu}_{ij} = \pm i\) [Fig.~\ref{fig:checkerboard}(a)(i)]. The ground state exhibits staggered \(\eta\)-pseudospin order with USP and CDM orders [Fig.~\ref{fig:checkerboard}(b)]. The BdG band structure shows an odd-parity \(\eta\)-pseudospin splitting [Fig.~\ref{fig:checkerboard}(c)(i)] with \(\eta^m = \pm1\) in the bands. From the BZ mapping of splitting energy \(E^\eta_{\mathbf{k}} = \sum_{n=1}^4 E_{n\mathbf{k}}\eta^m_{n\mathbf{k}}\) in the occupied bands [Fig.~\ref{fig:checkerboard}(d)(i)], we observe a \(p\)-wave structure under \([C^\eta_2 || \mathcal{\tilde{P}}]\). On the other hand, we find an even-parity \(\eta\)-ALM under a spin-dependent version of sublattice bonds \cite{das24prl} with \(\tilde{\nu}_{ij} = \pm \sigma^3\) [Fig.~\ref{fig:checkerboard}(a)(ii)]. In the staggered \(\eta\)-pseudospin order [Fig.~\ref{fig:checkerboard}(b)], the bands develop even-parity spin-\(\eta\)-locked splitting with \(\eta^m \sigma^3 = \pm 1\) [Fig.~\ref{fig:checkerboard}(c)(ii)]. The BZ mapping of splitting energy \(E^{s\eta}_{\mathbf{k}} = \sum_{n=1}^4 E_{n\mathbf{k}}\eta^m_{n\mathbf{k}}\sigma_{n\mathbf{k}}^3\) [Fig.~\ref{fig:checkerboard}(d)(ii)] shows a \(d\)-wave structure under \([C^\eta_2 || \tilde{\text{R}}_{\pi/2}]\) with Shiba-dual rotation \(\tilde{\text{R}}_{\pi/2}\).

Similar analysis is performed on the honeycomb lattice (Fig.~\ref{fig:honeycomb}). For odd parity, we consider the sublattice currents in the Haldane model \cite{haldane88prl} with \(\tilde{\nu}_{ij} = \pm i\) [Fig.~\ref{fig:honeycomb}(a)(i)]. Although the pairing ground state was known in the previous literature \cite{zhang17pra, dosanjos26ax}, the BdG band splitting has not been studied. In the staggered \(\eta\)-pseudospin order [Fig.~\ref{fig:honeycomb}(b)], the bands develop odd-parity \(\eta\)-pseudospin splitting [Fig.~\ref{fig:honeycomb}(c)(i)] with a \(f\)-wave structure [Fig.~\ref{fig:honeycomb}(d)(i)] under \([C^\eta_2 || \mathcal{\tilde{P}}]\) and \([C^\eta_2 || \tilde{\text{R}}_{\pi/3}]\). Note that this \(\eta\)-ALM can manifest topological gapped states, such as chiral superconductivity \cite{zhang17pra} and Chern insulator \cite{haldane88prl}. Meanwhile, even-parity \(\eta\)-ALM occurs under sublattice spin bonds with \(\tilde{\nu}_{ij} = 0, \pm\sigma^3\) [Fig.~\ref{fig:honeycomb}(a)(ii)]. The staggered \(\eta\)-pseudospin order [Fig.~\ref{fig:honeycomb}(b)] induces an even-parity spin-\(\eta\)-locked splitting [Fig.~\ref{fig:honeycomb}(c)(ii)]. This splitting has a \(d\)-wave structure [Fig.~\ref{fig:honeycomb}(d)(ii)] under \([C^\eta_2 || \tilde{\text{M}}]\), where \(\tilde{\text{M}}\) is a Shiba-dual mirror symmetry.

\textit{Discussion---}We have identified \(\eta\)-ALM and discovered its odd- and even-parity splittings in the BdG bands. This discovery is achieved by revisiting the Shiba duality of spin and \(\eta\)-pseudospin orders, where we established the correspondence of symmetry and band-structure properties. A natural question is whether the \(\eta\)-pseudospin splitting persists away from half filling. Note that doping the attractive model corresponds to adding a Zeeman field to the repulsive model \cite{scalettar89prl}. Accordingly, the ground state develops a canted \(\eta\)-ALM with uniform \(\eta^3\)- and staggered \(\eta^{1,2}\)-pseudospin orders. In addition to the \(s\)-wave \(\eta^3\)-splitting, the bands develop alternate \(\eta^{1, 2}\)- or \(\eta^{1, 2}\sigma^3\)-splitting that are absent in the pure Hubbard model. This alternate splitting corresponds to pairing orders and may lead to interesting phenomena in experiments. On the other hand, the concepts of Shiba duality and \(\eta\)-ALM may be generalized to the other ALM models \cite{smejkal22prx1, smejkal22prx2, roig24prb}, which is an interesting topic for future work.

For experiments, important future work includes search for realization, measurable signals, and applications. The cold-atom systems are a promising platform for the experimental realization. With the availability of Hubbard interaction \cite{mazurenko17n, gall21n, taie22n, xu25n} and various hopping terms, including sublattice currents \cite{jotzu14n}, the bipartite Hubbard models can be synthesized for \(\eta\)-ALM ground states. The band splitting can be observed by cold-atom ARPES methods with Raman spectroscopy \cite{dao07prl}, radio-frequency spectroscopy \cite{stewart08n}, or quantum gas microscopy \cite{brown20np}. On the other hand, the search in quantum materials may open applications in superconducting devices. Potential candidates include materials with chiral orbital currents \cite{zhang22n}, Floquet-driven systems \cite{huang26prl, zhu26prl, li26prl, liu26prb}, or Van der Waals heterostructures. The band splitting may be observed by ARPES or tunneling spectroscopy. Similar to ALM-superconductor junctions \cite{papaj23prb, sun23prb, ouassou23prl, beenakker23prb}, \(\eta\)-ALM may exhibit orientation-dependent Andreev reflection and Josephson effects. The junctions may also involve rotations between pairing and charge orders, thereby manifesting \(\eta\)-pseudospin torque beyond normal Josephson effects.

\textit{Acknowledgments---}The author thanks Joel Moore for discussions on implementation of Hartree-Fock-Bogoliubov theory, Marc Vila for previous collaboration on altermagnetism, and ChatGPT for discussion on materials in this work. This work was supported in part by the Gordon and Betty Moore Foundation through the Emergent Phenomena in Quantum Systems (EPiQS) Initiative during the author’s Moore Postdoctoral Fellowship at the University of California, Berkeley.




\bibliography{reference}

\clearpage
\onecolumngrid

\begin{center}{\large\bf
Supplemental Material for\\``Shiba duality and \(\eta\)-altermagnetism: Pairing and charge orders in bipartite attractive Hubbard models''
}\end{center}

\setcounter{secnumdepth}{3}
\setcounter{section}{0}
\setcounter{equation}{0}
\setcounter{figure}{0}
\renewcommand{\theequation}{S\arabic{equation}}
\renewcommand{\thefigure}{S\arabic{figure}}
\newcommand\Scite[1]{[S\citealp{#1}]}
\makeatletter \renewcommand\@biblabel[1]{[S#1]} \makeatother

\addtocontents{toc}{\string\tocdepth@restore}

\tableofcontents

\section{Hartree-Fock-Bogoliubov theory}

In this section, we introduce the Hartree-Fock-Bogoliubov (HFB) theory that is used to study the \(\eta\)-altermagnetism (\(\eta\)-ALM) in charge and pairing ordered ground states of interacting fermions. The formalism is a Bogoliubov-de Gennes (BdG) extension of the Hartree-Fock theory from the particle branch to both particle and hole branches. For a more pedagogical introduction to the formalism of Hartree-Fock theory, see Supplemental Material of Ref.~\cite{lin24axkagome}.

\subsection{General formalism}

We first discuss the general formalism of the HFB theory. Consider the general Hamiltonian of interacting fermions
\begin{equation}
    H = \sum_{ab}(\mathcal{H}_0)_{ab} c_a^\dagger c_b + \frac{1}{2} \sum_{abcd} U_{acdb} c_a^\dagger c_c^\dagger c_d c_b
\end{equation}
in an arbitrary basis set \(\{a\}\). Here \(c_a^{(\dagger)}\) annihilates (creates) a fermion in the state \(a\), \(\mathcal{H}_0\) represents the noninteracting Hamiltonian with chemical potential \(\mu\), and \(U\) is the interaction. Our interest lies in the mean-field BdG theory, where the Hilbert space is composed of fermionic Gaussian states \(\ket{\Psi}\) with nonzero particle-hole and particle-particle condensates \(\epvl{c^\dagger c^{(\dagger)}}\). Note that the BdG formalism manifests the particle and hole branches under an intrinsic particle-hole symmetry. Correspondingly, the state \(\ket{\Psi}\) can be represented by a BdG density matrix
\begin{equation}
    P
    = \begin{pmatrix}
        P^{00} & P^{01} \\
        P^{10} & P^{11}
    \end{pmatrix},
\end{equation}
where the diagonal and off-diagonal blocks have particle-hole and particle-particle condensates as their matrix elements, respectively
\begin{equation}
    P^{00}_{ab} = \epvl{c_b^\dagger c_a}, \quad
    P^{11}_{ab} = \epvl{c_b c_a^\dagger} - \delta_{ba}, \quad
    P^{01}_{ab} = \epvl{c_b c_a}, \quad
    P^{10}_{ab} = \epvl{c_b^\dagger c_a^\dagger}.
\end{equation}
This density matrix obeys the Hermiticity condition \(P^\dagger = P\), the block relations
\begin{equation}
    P^{00} = -(P^{11})^T,
    \quad
    P^{01} = -(P^{10})^T,
\end{equation}
and the projector condition
\begin{equation}
    (P + D_{11})^2 = P + D_{11}
\end{equation}
with the hole-branch identity \(D_{11} = \text{diag}(0^{00}, 1^{11})\).

Similar to the Hartree-Fock theory, the mean-field energy of the state \(\ket{\Psi}\) can be computed as
\begin{equation}
    E[P] = \frac{1}{2}\Tr\left[P\frac{1}{2}(\mathcal{H}_{\text{BdG}, 0} + \mathcal{H}_\text{HFB}[P])\right].
\end{equation}
Here the noninteracting Hamiltonian takes the BdG form
\begin{equation}
    \mathcal{H}_{\text{BdG}, 0}
    = \begin{pmatrix}
        \mathcal{H}_0 & 0 \\
        0 & -\mathcal{H}_0^T
    \end{pmatrix}.
\end{equation}
Meanwhile, the HFB Hamiltonian
\begin{equation}
    \mathcal{H}_\text{HFB}[P]
    = \begin{pmatrix}
        \mathcal{H}_\text{HFB}^{00}[P] & \mathcal{H}_\text{HFB}^{01}[P] \\
        \mathcal{H}_\text{HFB}^{10}[P] & \mathcal{H}_\text{HFB}^{11}[P]
    \end{pmatrix}
\end{equation}
has the blocks
\begin{equation}
    \begin{aligned}
        \mathcal{H}_\text{HFB}^{00}[P]_{ab} &= \mathcal{H}_{0, ab} + \sum_{cd} (U_{acdb} - U_{acbd}) P^{00}_{dc}, \\
        \mathcal{H}_\text{HFB}^{11}[P]_{ab} &= (-\mathcal{H}_0^T)_{ab} + \sum_{cd} (U_{bdca} - U_{bdac}) P^{11}_{dc}, \\
        \mathcal{H}_\text{HFB}^{01}[P]_{ab} &= \sum_{cd} U_{abcd} P^{01}_{dc}, \\
        \mathcal{H}_\text{HFB}^{10}[P]_{ab} &= \sum_{cd} U_{cdab} P^{10}_{dc}.
    \end{aligned}
\end{equation}
This effectively noninteracting Hamiltonian defines the mean-field theory of the interacting fermions
\begin{equation}
    H_\text{HFB}[P] = \frac{1}{2} \sum_{ab} \psi_a^\dagger \mathcal{H}_\text{HFB}[P]_{ab} \psi_b
\end{equation}
with the Nambu spinors
\begin{equation}
    \psi = \begin{pmatrix}
        c \\ c^\dagger
    \end{pmatrix}.
\end{equation}

In the HFB theory, the mean-field ground state \(\ket{\Psi_\text{GS}}\) satisfies the self-consistent equation
\begin{equation}
    [P_\text{GS} + D_{11}, H_\text{HFB}[P_\text{GS}]] = 0
\end{equation}
by fully occupying the negative-energy spectrum of the mean-field Hamiltonian \(H_\text{HFB}[P_\text{GS}]\). Our numerical computation solves this ground state through an iterative algorithm. Starting from a random or designed density matrix \(P_0\), each iteration derives the HFB Hamiltonian \(H_\text{HFB}[P]\) and updates the density matrix \(P\) with the negative-energy spectrum. The ground-state density matrix \(P_\text{GS}\) is determined when energy \(E[P]\) and density-matrix elements reach the desired convergence.

\subsection{Symmetry-breaking orders}

For a ground state \(\ket{\Psi_\text{GS}}\) with density matrix \(P_\text{GS}\), we can compute its charge, spin, and pairing orders to understand its symmetry breaking. For the lattice models in our study, we consider the basis set \(a = i\sigma\) with the indices of site \(i\) and spin \(\sigma\). We are interested in the onsite charge, spin, and pairing orders in the ground state \(\ket{\Psi_\text{GS}}\). The onsite charge and spin orders can be computed from the particle-branch block
\begin{equation}
    s_i^\mu
    = \Epvl{c_i^\dagger \frac{\sigma^\mu}{2} c_i}
    = \Tr\left(P^{00}_i\frac{\sigma^\mu}{2}\right).
\end{equation}
Here \(s_i^\mu = (s^0, \mathbf{s})\) are the onsite charge and spin orders at site \(i\), \(P^{00}_i\) is the \(2\times2\) onsite particle-branch density matrix at site \(i\), and \(\sigma^\mu = (\sigma^0, \boldsymbol{\sigma})\) with \(\mathbf{\sigma} = (\sigma^1, \sigma^2, \sigma^3)\) are the \(2\times2\) identity matrix and Pauli matrices. Meanwhile, the onsite pairing order can be computed from the particle-particle condensate
\begin{equation}
    p_i^\mu
    = \epvl{c_i^\dagger \frac{\sigma^\mu}{2} (i\sigma^2) (c_i^\dagger)^T}
    = \Tr\left[P^{10}_i \frac{\sigma^\mu}{2} (i\sigma^2)\right].
\end{equation}
The pairing orders \(p^0_i\) and \(\mathbf{p}_i\) correspond to the onsite singlet and triplet pairings, respectively. Note that the onsite triplet pairings are strictly zero \(\mathbf{p}_i = \mathbf{0}\) due to Fermi statistics.

\subsection{Band-structure analysis}

To study the \(\eta\)-pseudospin splitting in \(\eta\)-ALM, we need to compute the band structure in momentum space. This computation is achieved by Fourier transforming the real-space Hamiltonian to momentum space. Consider a noninteracting real-space Hamiltonian in the BdG formalism
\begin{equation}
    \tilde{H}_\text{BdG} = \frac{1}{2} \sum_{\tilde{i}\tilde{i}'\tilde{\tau}\tilde{\tau}'} \psi_{\tilde{i}\tilde{\tau}}^\dagger (\tilde{\mathcal{H}}_\text{BdG})_{\tilde{i}\tilde{i}'\tilde{\tau}\tilde{\tau}'} \psi_{\tilde{i}'\tilde{\tau}'},
\end{equation}
which represents either the noninteracting Hamiltonian or the mean-field HFB Hamiltonian. Here \(\tilde{i}\) and \(\tilde{\tau}\) are the indices of Bravais lattice site and sublattice under the periodicity of the Hamiltonian. For the Nambu spinor \(\psi_{\tilde{i}\tilde{\tau}}\), we adopt the convention with the time-reversal spin flip \(i\sigma^2\) in the hole branch
\begin{equation}
    \psi_{\tilde{i}\tilde{\tau}} = \begin{pmatrix}
        c_{\tilde{i}\tilde{\tau}} \\ (i\sigma^2) (c_{\tilde{i}\tilde{\tau}}^\dagger)^T
    \end{pmatrix}.
\end{equation}
The Fourier transform is performed as
\begin{equation}
    \psi_{\tilde{i}\tilde{\tau}} = \frac{1}{\tilde{N}_\text{BL}^{1/2}} \sum_{\mathbf{k}} \psi_{\mathbf{k}\tilde{\tau}} e^{i \mathbf{k} \cdot \mathbf{r}_{\tilde{i}\tilde{\tau}}}, \quad
    \psi_{\mathbf{k}\tilde{\tau}} = \begin{pmatrix}
        c_{\mathbf{k}\tilde{\tau}} \\ (i\sigma^2) (c_{-\mathbf{k}\tilde{\tau}}^\dagger)^T
    \end{pmatrix},
\end{equation}
where \(\tilde{N}_\text{BL}\) is the number of Bravais lattice sites. This transformation yields the momentum-space Hamiltonian
\begin{equation}
    \tilde{H}_\text{BdG} = \frac{1}{2} \sum_{\mathbf{k}\tilde{\tau}\tilde{\tau}'} \psi_{\mathbf{k}\tilde{\tau}}^\dagger (\tilde{\mathcal{H}}_\text{BdG})_{\mathbf{k}\tilde{\tau}\tilde{\tau}'} \psi_{\mathbf{k}\tilde{\tau}'},
\end{equation}
where the momentum-space BdG Hamiltonian reads
\begin{equation}
    (\tilde{\mathcal{H}}_\text{BdG})_{\mathbf{k}\tilde{\tau}\tilde{\tau}'} = \sum_{\tilde{i}'} (\tilde{\mathcal{H}}_\text{BdG})_{\tilde{0}\tilde{i}'\tilde{\tau}\tilde{\tau}'} e^{-i \mathbf{k} \cdot (\mathbf{r}_{\tilde{0}\tilde{\tau}} - \mathbf{r}_{\tilde{i}'\tilde{\tau}'})}.
\end{equation}
The BdG band structure is then obtained by diagonalizing the BdG Hamiltonian, where the BdG states \(\ket{u_{n\mathbf{k}}}\) and their dispersion energies \(E_{n\mathbf{k}}\) are determined.

\end{document}